 \documentstyle[twocolumn,prl,aps,psfig,amssymb]{revtex}
\begin{document}
 \twocolumn[\hsize\textwidth\columnwidth\hsize\csname @twocolumnfalse\endcsname
\draft
\begin{title}
 {Addition Spectra of Chaotic Quantum Dots:\\
 Interplay between Interactions and Geometry }
\end{title}
\author{
    Kang-Hun Ahn$^{1}$, Klaus Richter$^{1}$, and In-Ho Lee$^{2}$}
\address{$^{1}$Max-Planck-Institut f\"ur Physik komplexer Systeme, 
N\"othnitzer Str. 38, 01187 Dresden, Germany      \\ }
\address{$^{2}$School of Physics, Korea Institute for Advanced Study,
Cheongryangri-dong, Seoul 130-012, Korea}
\date{\today}
\maketitle

\widetext

\begin{abstract}
We investigate the influence of interactions and geometry on
ground states of clean chaotic quantum dots using the self-consistent
Hartree-Fock method. 
We find two distinct regimes of interaction strength:  While
capacitive energy fluctuations $\delta \chi$ follow
approximately a random matrix prediction for weak 
interactions, there is a crossover to a regime
 where $\delta \chi$ is strongly enhanced and
scales roughly with interaction strength.
This enhancement is related to the rearrangement of
charges into 
ordered states near the quantum dot edge.
This effect is non-universal depending on dot shape and size. 
It may provide additional insight into
recent experiments on statistics of Coulomb blockade peak
spacings.
\end{abstract}
\pacs{05.45.+b,72.10Fk,73.20.Dx}
]
\narrowtext

The understanding of the interplay between many-body interactions and chaos 
has evolved to a prominent field in mesoscopic physics. This challenging 
issue brings together two seemingly disconnected 
fields, namely many-particle physics and quantum chaos.
Classically chaotic independent-particle dynamics is found in mesoscopics
for both, disordered systems due to scattering at impurities and
{\it clean} quantum dots, where the shape of the confining 
potential can give rise to chaotic electron motion. 
While 
interaction effects in disordered systems have been intensively studied in the 
recent past\cite{diffusive}, accounts on the inter-relation between many-body
effects and chaotic dynamics in ballistic quantum dots
are still rare (see, e.g., Refs.~\cite{stopa,magnetism}). 
The latter systems can be modelled by quantum billiards which have served as 
prototypes to investigate quantum signatures
of integrable and chaotic single-particle dynamics\cite{LesHqchaos}. Hence a 
generalization of these quantum chaos concepts to interacting particles in 
billiards appears natural. 
In this letter we approach this problem within a self-consistent 
Hartree-Fock (SCHF) approximation 
after checking its validity by comparing with exact diagonalization.

Our work was partly motivated by new measurements of charge transport through 
quantum dots with variable size and shape, which represent ideal tools 
to investigate experimentally interaction effects in confined chaotic systems.
If such a dot with $N$ electrons and ground state energy $E_N$ is only 
weakly coupled to external leads a conductance peak
is observed, upon tuning a gate voltage connected
to the dot, whenever the chemical potential $\mu_{N}=E_{N}-E_{N-1}$
 coincides with that of the leads.
In between the pronounced conductance peaks the current through the dot
is suppressed due to Coulomb blockade\cite{kastner92}.
The spacing between neighboring peaks as function of 
gate voltage is proportional to the capacitive energy 
\begin{equation}
\chi_{N} = E_{N+1} - 2E_N + E_{N-1} \; .
\label{chiN}
\end{equation}
A number of recent experiments\cite{sivan,simmel,patel,simmel2} showed
that fluctuations of $\chi_{N}$ resemble a Gaussian distribution, while 
the assumption of a constant interaction (CI), which properly describes the 
mean peak spacing, combined with random matrix theory (RMT), predicts a
Wigner-Dyson distribution.
Two reasons have been proposed for the Gaussian
distribution, the scrambling of the single-particle spectrum due to 
interactions\cite{sivan,cohen,levit,bonci,walker} 
and the deformation of the dot by changing the gate voltage\cite{vallejos}. 

However, it is not yet understood why the measured  peak spacing 
fluctuations,
$\delta \chi = (\langle \chi_{N}^2\rangle -\langle \chi_{N}\rangle^2)^{1/2}$, 
significantly vary between the different experiments:
Patel et al.\cite{patel} found rather small fluctuations which are
comparable to the single-particle mean level spacing $\Delta$.
 They are in line with the RMT+CI model predicting
$ \delta \chi =\Delta (4/\pi -1)^{1/2} \simeq 0.52 \Delta$,
and with RPA calculations for weak interactions (high densities,
$r_{s} < 1$)\cite{blanter,berkovits}. On the contrary, the other
experiments yield large fluctuations which scale rather
with $\langle \chi_{N}\rangle$ than with $\Delta$:
$\delta \chi \approx 0.06 - 0.15\langle \chi_{N}\rangle$ in
Refs.~\cite{sivan,simmel,simmel2}.
E.g., in the recent experiment \cite{simmel2},
where the shape of the dot is expected
not to be distorted by tuning the gate voltage, $\delta \chi$ was
found to be 
15 times larger than predicted by RMT.

Thus two main open questions arise:
(i) what are the responsible mechanisms for the large fluctuations
and (ii) why do they show such a sample-dependent behavior, assuming that
the samples are chaotic?

Here we consider clean chaotic quantum dots
and study the interplay between the Coulomb interaction of the electrons
and the geometry of the dots.
We propose a mechanism for enhanced capacitive energy 
fluctuations $\delta \chi$ in chaotic structures and show that:

(i) interaction drives the systems to a regime of large, 
non-universal fluctuations by developing ordered charge states near 
the edges.

(ii) the shape of the confinement is an important factor for the
formation of such states and hence for the fluctuations. 

To our knowledge, all related theoretical studies consider interaction 
effects for disordered models\cite{sivan,cohen,levit,bonci,walker},
apart from the work by Stopa\cite{stopa}, who suggested 
the occurence of strongly scarred states as a mechanism for enhanced 
fluctuations.
Our calculations show that
$\delta \chi$ can be enhanced in the absence of scarred states.

\begin{figure}
\centerline{\psfig{figure=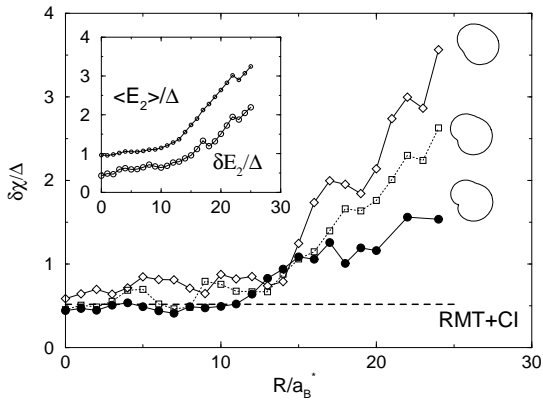,width=0.4\textwidth}}
\caption{ 
Fluctuations  of capacitive energies
$\delta \chi = (\langle \chi_{N}^2\rangle $ $ -\langle 
\chi_{N}\rangle^2)^{1/2}$ 
as a function
of interaction strength for interacting electrons in three chaotic billiards
of different shape shown on the right 
(defined 
with $\delta=1.5$, $b\!=\!c\!=\!0.12,0.16$, and 0.2 from above, see the text).
$\delta \chi$ shows a clear cross-over 
near $R/a_{B}^{*}\!=\!12$ and increases 
with increasing  $R/a_{B}^{*}$ in the strong interaction regime.
An analysis of 
the statistical error in computing $\delta \chi$ shows that it
is of the order of the fluctuations visible in the curves.
Inset: The mean and rms value of 
$E_2\!=\! \epsilon_{N+2}^{(N)}\!-\! \epsilon_{N+1}^{(N)} $ for the
quantum dot of the bottom right inset.
Note that for larger interactions $\langle E_2 \rangle \neq \Delta$.}
\label{xn_debi}
\end{figure}

We model the quantum dots by a family of billiards which arise
from a deformation of the disk\cite{robnik}. They have  
already been employed as non-interacting models
in the context of Coulomb blockade\cite{vallejos,bruus}.
Here they will serve as 
appropriate tools to study 
interaction  effects
upon changing the billiard geometry.
The confinement is defined by  a hardwall 
potential $U$ satisfying $U(u,v)=0$ inside a
domain $D$ and $U(u,v)=\infty$ outside.
Using complex coordinates $\omega=u+iv$ and $z=x+iy$, 
$D$ is defined by the conformal mapping\cite{robnik}
$\omega(z)=R(z +bz^{2}+ce^{i\delta}z^{3})/(\sqrt{1+2b^2+3c^2})  ;
   |z|< 1 \; $.
The real parameters $b, c$, and $\delta$ determine the shape,
and $R$ defines the dot size and area, $A\!=\!\pi R^2$.
For the deformations to be considered (insets in Fig.~\ref{xn_debi})
the classical single-particle dynamics is predominantly chaotic 
(described by a large Kolmogorov entropy
 for the strongest deformation\cite{bruus}),
 and we found that the single-particle 
energies exhibit Wigner-Dyson statistics.

To include interactions we first calculate the single-particle
eigenfunctions $\phi_i$ and -energies $\epsilon_i$ of a deformed billiard. We then
use them to construct the Coulomb interaction matrix elements
$ V_{ijkl}=\int {\rm d}{\bf r}_1d{\bf r}_2
\phi_{i} ({\bf r}_1) \phi_j ({\bf r}_2)
V({\bf r}_1-{\bf r}_2)
\phi_k ({\bf r}_1) \phi_l ({\bf r}_2)$
with  $V({\bf r}) \!=\!e^{2}/\epsilon r$,
and the many-electron Hamiltonian 
\begin{equation}
H=\sum_{i}\epsilon_{i} c_{i}^{\dagger}c_{i} +
\frac{1}{2} 
\sum_{ijkl}V_{ijkl} c_{i}^{\dagger} c_{j}^{\dagger} c_{l} c_{k} \; .
\label{hamiltonian}
\end{equation}
Here 
$c_{i}^{\dagger}$($c_{i}$) creates (annihilates) the i-th 
eigenstate of the non-interacting Hamiltonian.
In terms of energy units ${\hbar^2}/{(2m^*R^{2})} $,
$V_{ijkl}$ is proportional to the  dimensionless interaction strength,
or system size, $R/ a_{B}^{*}$, where 
$a_{B}^*=\hbar^{2}\epsilon/m^{*}e^{2}$ is the effective Bohr radius.
Eq.~(\ref{hamiltonian}) can serve as a starting point for both
exact diagonalization for few electrons\cite{ahn} and 
SCHF calculations. 

\begin{figure}
\centerline{\psfig{figure=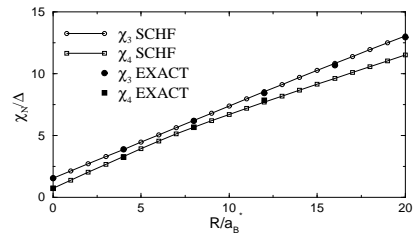,width=0.3\textwidth}}
\caption{ 
Comparison of self-consistent Hartree-Fock (SCHF) and exact diagonalization
(EXACT) results of capacitive energies $\protect \chi_N$, Eq.~(\ref{chiN}), 
as a function of interaction strength
for $N\!=\!3$ (upper) and $N\!=\!4$ (lower curve) for the billiard 
geometry shown in the bottom inset of Fig.~\ref{xn_debi}. 
}
\label{fig:SCHF_EXACT}
\end{figure}

In this work we focus on {\em ground state} properties for many-electron 
quantum dots and employ SCHF systematically for systems up to 27 spinless electrons.
The SCHF allows us to calculate the ground state energies even in the presence
of two-electron density correlations which are ignored
in the RPA perturbation calculations \cite{blanter,berkovits}.
To check the validity of our SCHF results for the capacitive energies $\chi_N$
we first compare them with exact diagonalization results
for few electrons. As shown in Fig.~\ref{fig:SCHF_EXACT} the 
SCHF results are very accurate even for strong interactions.

The SCHF results for $\chi_N$ do not show any visible dependence 
on $N$ but rather on the interaction strength (system size) $R/a_B^*$.
Let us thus consider the mean fluctuations $\delta \chi$, where
the statistical average $\langle \ldots \rangle $ is performed
over $N$ as in the experiments\cite{sivan,simmel,patel,simmel2}.

Our main results are summarized in Fig.~\ref{xn_debi}. It 
shows $\delta \chi$ as a function of interaction strength
for three different geometries depicted as insets close to the curves.
For small size quantum dots, $R/a_B^*  \stackrel{<}{\sim} 12 $,
$\delta \chi$  is close to the RMT prediction.
Most interestingly, we find for larger  $R/a_B^*$
a {\it cross-over} to a regime where the capacitive energy fluctuations
increase roughly linearly with increasing interaction strength.
To our knowledge such a clear-cut cross-over behavior
has not been found in any tight-binding model for disordered systems.
This indicates that it could be a property of the charge distribution
in confined ballistic systems.

Using the standard definition of $r_s$, namely
$r_s^2 = A/(\pi N {a_B^*}^2)$, 
we have $r_{s} = (R/a_B^*)/\sqrt{N}$.
Since we calculate $\delta \chi$ over a range $3 \le N  \le 27$ for
a given value of $R/a_B^*$ in Fig.~\ref{xn_debi}, $r_{s}$ has not
a well-defined value but is in the range
$(1/\sqrt{27})R/a_{B}^{*} \leq r_{s} \leq (1/\sqrt{3})R/a_{B}^{*}$.
Hence the cross-over value of $R/a_{B}^{*}$ in Fig.~\ref{xn_debi} 
implies that $\delta \chi$ retains its RMT value up to $r_s \approx 2$ 
which is consistent with the RPA approach of Ref.~\cite{blanter}.

In the following we discuss the mechanism responsible for
the cross-over to enhanced fluctuations $\delta \chi$. To this end 
it  proves convenient to decompose $\chi_{N+1}$ as\cite{blanter}
\begin{equation}
\chi_{N+1} \! \simeq  \!
 \left[ \epsilon_{N+2}^{(N+1)}\! - \! \epsilon_{N+2}^{(N)}\right] +
 \left[ \epsilon_{N+2}^{(N)}\! - \!\epsilon_{N+1}^{(N)}\right]
\!\equiv\! E_1\! + \! E_2.
\label{koopmans}
\end{equation}
Here, $\epsilon_{j}^{(i)}$ denotes the $j$-th eigenvalue of the
SCHF Hamiltonian 
with $i$ electrons. Quantitatively,
Eq.~(\ref{koopmans}), which relies on Koopmans' approximation, 
does not give the same value as $\delta \chi$ in Fig.~\ref{xn_debi}.
However, Eq.~(\ref{koopmans}) shows the same qualitative behavior 
with the same cross-over point but slightly larger slope than in the SCHF case,
Fig.~\ref{xn_debi}. 

The quantity $E_{1}$ in Eq.~(\ref{koopmans}), which describes
the interaction  energy 
between the electrons in
  $(N\!+\!1)$th and $(N\!+\!2)$th state,
has been considered
 in Ref.~\cite{blanter}.
$E_{2}$ is the energy difference between two adjacent single-particle 
levels of the same Hamiltonian.
In the RPA approach\cite{blanter},
it was claimed that $\langle E_2 \rangle$
is given by $\Delta$ and its fluctuations $\delta E_2$
follow the RMT value $\simeq 0.52\Delta$.
As displayed in the inset of Fig.~\ref{xn_debi}
our calculations show that this seems to be true for
$R/a_B^{*}\stackrel{<}{\sim}  12 $, which includes the
regime $r_s\! <\!1$, where RPA is reliable. 
However, $\langle E_2 \rangle/\Delta$ grows as the system size increases,
and $\delta E_{2} /\Delta$ and $\delta E_{2} /\langle E_2 \rangle$
do no longer follow the RMT prediction for $R/a_B^*   \gtrsim 12$.

The similarity of the cross-over behavior of $\delta \chi$ 
and $\delta E_{2}$ in  Fig.~\ref{xn_debi}  suggests that
an analysis of $E_2$ proves appropriate
 to understand
the enhanced fluctuations $\delta \chi$.

\begin{figure}
\centerline{\psfig{figure=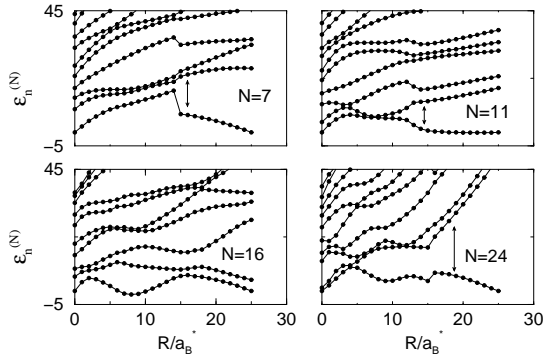,width=0.4\textwidth}}
\caption{Unoccupied SCHF levels $\epsilon_n^{(N)}, n\!=\!N+\!1,
N\!+\!2, \ldots$ as a function of interaction strength 
for a chaotic quantum dot (bottom right inset in Fig.~\ref{xn_debi})
with $N\!=$7, 11, 16, and 24 electrons.
Note the sudden jumps in the level spacings $E_2$ between the 
lowest two adjacent levels 
which are caused by abrupt changes in the mean-field potential.
For clarity, the levels in each panel are shifted by an amount 
$a(N)+b(N) R/a_{B}^{*}$.
The energy units are $\hbar^{2}/2m^{*}R^{2}$.}
\label{uhfn7}
\end{figure}

What happens in the regime $R/a_B^* \gtrsim 12$?  
To answer this question we present in Fig.~\ref{uhfn7} the evolution
of unoccupied SCHF single-particle levels $\epsilon_n^{(N)}, n\!=\!N\!+\!1, 
N\!+\!2, \ldots $
with increasing interaction for a few representative values of $N$.
We find that the enhancement of mean and rms values of
$E_2 \!=\! (\epsilon_{N+2}^{(N)}- \epsilon_{N+1}^{(N)})$ is
accompanied by the appearance of sudden jumps
of the single particle levels as marked by vertical arrows in Fig.~\ref{uhfn7}.
(Note however that not for all values of $N$ 
the energy levels show such jumps in the $R/a_B^*$ range considered,
e.g.\ $N\!=\!16$.)

\begin{figure}
\centerline{\psfig{figure=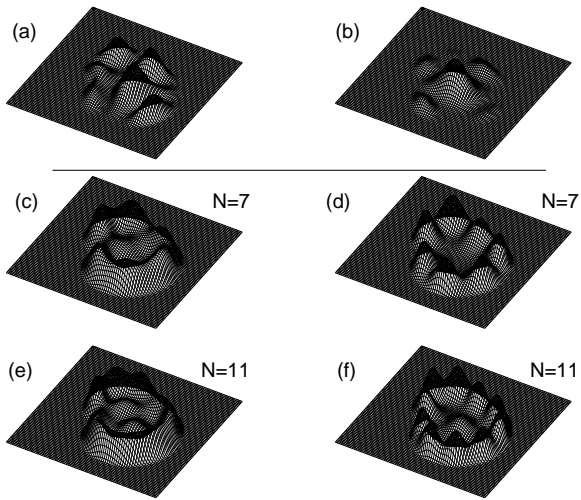,width=0.45\textwidth}}
\caption{Charge reordering with increasing interaction 
for a chaotic dot (bottom inset in Fig.~\ref{xn_debi});
$R/a_{B}^* = 5$ (15) in left (right) panels.
Panels (a) and (b) show the probability density of the
first unoccupied SCHF single-particle state $n=8, N=7$ (lowest
curve in upper left panel, Fig.~\ref{uhfn7}). In
(c) to (f) the total charge densities for $N=7,11$ are displayed.}
\label{totdensity}
\end{figure}

The abrupt changes in the SCHF single-particle energies  go along with a
rearrangement of the corresponding wavefunctions as shown in  
Fig.~\ref{totdensity}(a,b). This 
indicates a change in the total charge densities and thereby in
the underlying mean-field potential.
The two lower left (right) panels in Fig.~\ref{totdensity} depict
the total charge densities for interaction strength  $R/a_B^* = 5$ (15)
for two selected, representative $N$. 
A systematic analysis of a large number of geometries and
particle numbers allowed us to extract two common trends
visualized in the figure:

(i) With growing system size the electrons tend to be 
localized near the edge of the dot. {\em  For $R/a_B^* \gtrsim 12$ the
charges are reorganized in an ordered structure similar
to a one-dimensional crystal or a charge density wave state.}

(ii) An abrupt change in the mean-field potential for fixed $N$ 
occurs when an electron from the inner region of the dot moves into a 
state close to the edge.

Let us discuss the connection to the increase in $\langle E_2\rangle$ 
and $\delta E_2$. Electrons in states extending over the whole 
dot for small $R/a_B^*$ move with increasing interaction to 
the edge due to the long-range Coulomb repulsion.
When the states near the edge are fully occupied 
(see e.g.\ Fig.~\ref{totdensity}(d))
the corresponding mean field reduces the accessible space for the next 
electron above the Fermi level (crossover from \ref{totdensity}(a) to (b)). 
The reduced effective size of the dot gives rise to an enhanced level 
spacing $\langle E_2\rangle$ which is accompanied by enhanced fluctuations.
Note that $\delta E_2/\langle E_2\rangle$ also deviates from the RMT
value of $\approx 0.52$. This indicates that 
even for a chaotic geometry {\em the SCHF single-particle states near
the Fermi level in the strong interaction regime do no longer follow RMT 
due to charge rearrangement near the edge.}

The reorganized charge states at the edge differ from a two-dimensional 
Wigner crystal 
where the electrons are localized due to
electron correlation energy\cite{koulakov2}.

One can distinguish two kinds of ordered states:

For {\em  smaller} $N$ a phase which might 
be called a ``crystal chain'' evolves 
where the electrons are localized near the edge with
clear spatial separation (see, e.g., Fig.~\ref{totdensity}(d)).
Due to quantum fluctuations of such a 1D crystal lattice
this phase will be bounded from above by some critical electron number $N_{c}$.
Since the minimum distance between the electrons should be larger than a
Fermi wavelength $\lambda_F$, the maximum number of electrons  
which can reside at the edge is $ \approx L_p/\lambda_F$,
where $L_p$ is the perimeter of the dot.
Since in the crystal chain state all electrons are at the edge,
$N_{c}$ can be estimated as
$ N_{c}\approx  L_p^2 / A$, 
where we used $\lambda_F \approx  R/\sqrt{N}$. For the dot geometries
used
 we find $N_{c}$ to be of order of 10.
This rough estimation is in line with our numerical results where
for $N=7$ all electrons are near the edge, whereas for $N=11$ only 10 
electrons seem close to the edge with one electron near the 
center of the dot (see Fig.~\ref{totdensity}(f)).

For {\em larger} electron number $N\! >\! N_c$, as in Fig.~\ref{totdensity}(f), 
the charge density exhibits a modulation close to the edge rather than well 
separated charges. 
Such ordered, charge density wave like structures
\cite{hirose} persist up to high electron numbers.
We still found these states for $N\!=\!40$, the highest $N$ considered.
They play the same role for the enhancement of $\delta \chi$ 
(see, e.g., Fig.~\ref{uhfn7} for $N\!=\!24$) as the crystal type states do for
$N\! <\! N_c$.

A few further remarks are due:

(i) 
A strongly deformed quantum dot where part of the boundary
is concave\cite{kink} (insets in Fig.~\ref{xn_debi})
renders the formation of edge states more difficult than the dot
with a shape close to a disk due to the interplay between the
long range interaction and geometry.
This  is the reason why the fluctuations 
$\delta \chi$ are the largest in the latter case and turn out to be 
system dependent, even if the non-interacting geometries all exhibit
chaotic dynamics.

(ii)
The accumulation  of charge near the edge alone 
is not sufficient to
produce enhanced fluctuations $\delta \chi$, as visible in 
Fig.~\ref{totdensity}(c, e) for weakly interacting dots ($R/a_{B}^*$=5).
This weak interacting case, where $\delta \chi$ is still similar to the 
RMT value, can be understood within the RPA approach of Ref.\cite{blanter}, 
where the corresponding contribution to $\delta \chi$ was
shown to be of order of $\Delta$ or smaller, independent of the 
interaction strength.
The formation of ordered states, which no longer follow RMT predicitions,
is necessary to generate enhanced fluctuations.

(iii) The shape dependence of the capacitive energy fluctuations
found within our model, which does not include effects, e.g.\  from the 
dot surroundings, temperature, and spin, 
may be considered as one possible mechanism
which can lead to the different widths in the peak spacing distributions
of the various experiments [6-9].

An experimental observation of the interaction dependent 
crossover of $\delta \chi$ would be particularly interesting.
For GaAs samples with $a_B^*\approx 10$ nm, the cross-over 
may appear for a system size close to $0.1\mu$m.
Recent experiments\cite{austing} on small quantum dots
without spatial symmetry may be a good testing ground.

To conclude, 
we considered interaction effects in clean chaotic quantum dots.
We propose as a mechanism for enhanced capacitive energy
fluctuations the formation of ordered charge states near
the quantum dot edges.
In the regime where the ordered states are developed
the fluctuations are geometry-dependent, even for chaotic systems.

We thank R. Berkovits, H.\ Castella, A.\ Cohen, Y. Gefen, 
S.\ Kettemann, A.D. Mirlin, G.\ Montambaux, and J.-L. Pichard
for helpful discussions. KHA thanks M. Takahashi for useful comments on 
numerical calculations.


\begin{references}
\bibitem{kink} 
        The shape sensitivity of the fluctuations $\delta \chi$ is not 
        specific to the ``kink'' in the structure shown in the bottom inset 
        of Fig.~\ref{xn_debi}, but also exists in deformed geometries with 
        smooth confinement.
\end{references}
\end{document}